\documentclass[epj,nopacs,final]{svjour}
\usepackage[dvips]{graphicx}
\begin{document}
\title{
\vspace*{-2.0\baselineskip}
\rightline{\textrm{\normalsize\rm CERN-TH/2003-254}} \\[3mm]
Improved $\eta^\prime$-Meson Distribution Amplitudes from 
Inclusive $\Upsilon (1S) \to \eta^\prime X$ Decay 
} 
%
%
\author{Ahmed Ali\inst{1}\thanks{
   On leave of absence from 
        Deutsches Elektronen-Syn\-chro\-tron 
        DESY, Hamburg, FRG.}%
   \and Alexander Ya. Parkhomenko\inst{2}\thanks{
   On leave of absence from
        Department of Theoretical Physics,
        Yaroslavl State University,
        Sovietskaya~14, 150000 Yaroslavl, Russia.}%
%
}                     
%
%
\institute{ 
  Theory Division, CERN, CH-1211 Geneva 23, Switzerland 
  \and 
  Institut f$\ddot {\rm u}$r Theoretische Physik,
  Universit$\ddot {\rm a}$t Bern, CH-3012 Bern, Switzerland
}
%
\date{\phantom{Received: date / Revised version: date}}
%
\abstract{
We calculate the $\eta^\prime$-meson energy spectrum in the 
$\Upsilon (1S) \to \eta^\prime g g g \to \eta^\prime X$ decay 
in the leading-order perturbative QCD in the static-quark limit 
for the orthoquarkonium. Our principal result is the extraction 
of parameters of the $\eta^\prime g^* g$ effective vertex function 
(EVF) involving a virtual and a real gluon from the available data 
on the hard part of the $\eta^\prime$-meson energy spectrum. The 
perturbative-QCD based framework provides a good description of 
the available CLEO data, allowing one to constrain the lowest 
Gegenbauer coefficients $B^{(q)}_2$ and $B^{(g)}_2$ of the 
quark-antiquark and gluonic distribution amplitudes of the 
$\eta^\prime$-meson. The resulting constraints are combined with 
the existing ones on these coefficients from an analysis of the 
$\eta^\prime - \gamma$ transition form factor and the requirement
of positivity of the EVF, yielding $B^{(q)}_2 (\mu_0^2) = -0.008 
\pm 0.054$ and $B^{(g)}_2 (\mu_0^2) = 4.6 \pm 2.5$ for $\mu_0^2 =
2$~GeV$^2$. This reduces significantly the current uncertainty 
on these coefficients. 
\PACS{
      {PACS-key}{discribing text of that key}   \and
      {PACS-key}{discribing text of that key}
     } 
} 
\titlerunning{
Improved $\eta^\prime$-meson distribution amplitudes 
from inclusive $\Upsilon (1S) \to \eta^\prime X$ decay
}
\authorrunning{A.~Ali, A.Ya.~Parkhomenko}
\maketitle
\section{Introduction}
\label{sec:introduction}

A quantitative description of the rare decays with the
$\eta^\prime$-meson production, such as $B \to \eta^\prime K^{(*)}$,   
$B \to \eta^\prime X_s$, $\Upsilon (1S) \to \eta^\prime X$, 
requires an understanding of the $\eta^\prime g^* g^{(*)}$ effective 
vertex function (EVF), $F_{\eta^\prime g^* g^{(*)}} (q_1^2, q_2^2,
m_{\eta^\prime}^2)$ [also called the $\eta^\prime - g$ 
transition form factor when one of the gluons is on the mass shell].  
If one or both of the gluons entering into the EVF in such decays
are far from their mass shell, the dependence of the EVF on the gluon 
virtualities should be included in the theoretical analysis of decays. 
For energetic $\eta^\prime$-meson, the QCD hard-scattering 
approach can be used to derive the required 
EVF~\cite{Muta:1999tc,Ali:2000ci,Kroll:2002nt,Ali:2003kg}. As the 
$\eta^\prime$-meson has a relatively large mass, $m_{\eta^\prime} = 
958$~MeV, its effect should be taken into account in applications 
of the $\eta^\prime g^* g^{(*)}$ EVF to physical processes, in 
particular when the gluon virtualities are time-like~\cite{Ali:2003kg}. 
Moreover, the inclusion of the $\eta^\prime$-meson mass results 
in a pole-like form for the $\eta^\prime - g$ transition form factor:   
\begin{equation}
F_{\eta^\prime g} (p^2) \equiv 
F_{\eta^\prime g^* g} (p^2, 0, m_{\eta^\prime}^2) = 
\frac{m_{\eta^\prime}^2 \, H (p^2)}{p^2 - m_{\eta^\prime}^2},
\label{eq:EVF-PhF}
\end{equation}
This form was introduced by Kagan and Petrov~\cite{Kagan:1997qn}. 
These authors also suggested to ignore the dependence of the
function~$H (p^2)$ on the gluon virtuality and approximate it 
by a constant value, $H_0 = 1.7$~GeV$^{-1}$, resulting from the 
analysis of the $J/\psi \to \eta^\prime \gamma$ 
decay~\cite{Atwood:1997bn}. Also it was subsequently shown that 
the hard part of the $\eta^\prime$-meson energy spectrum in the 
inclusive $\Upsilon (1S) \to \eta^\prime X$ decay~\cite{Kagan:2002dq} 
is in a qualitative agreement with the spectrum measured recently 
by the CLEO collaboration~\cite{Artuso:2002px}. Starting form 
the above observations, a quantitative analysis of the 
$\eta^\prime$-meson energy spectrum in this decay was undertaken 
by us~\cite{Ali:2003vw} and its results are briefly summarized 
in this report.

\section{The $\eta^\prime$-Meson Wave-Function}
\label{sec:wave-function}

In the quark-mixing scheme, the Fock-state decomposition of 
the $\eta^\prime$-meson wave-function is as follows:  
\begin{equation}
|\eta^\prime \rangle = \sin \phi \, |\eta^\prime_q \rangle + 
\cos \phi \, |\eta^\prime_s \rangle + |\eta^\prime_g \rangle , 
\label{eq:WF-decomposition} 
\end{equation}
where 
$|\eta^\prime_q \rangle \sim |\bar u u + \bar d d \rangle / \sqrt{2}$, 
$|\eta^\prime_s \rangle \sim |\bar s s \rangle$, and 
$|\eta^\prime_g \rangle \sim | gg \rangle$ are the light-quark, 
strange-quark and gluonic components, respectively.  
For an energetic $\eta^\prime$-meson in a process, its 
wave-function can be described in terms of the quark-antiquark, 
$\phi^{(q)}_{\eta^\prime} (x, Q^2)$, and gluonic,
$\phi^{(g)}_{\eta^\prime} (x, Q^2)$, light-cone distribution 
amplitudes (LCDAs).    
In the LCDAs above, $x$~is the momentum fraction of one of the 
partons inside the meson and $Q^2$ is a typical hard scale of 
the process. Note that these LCDAs mix under the scale evolution.   

%
%
\begin{figure}[tb]
\centerline{
\resizebox{0.55\hsize}{!}{
\includegraphics*{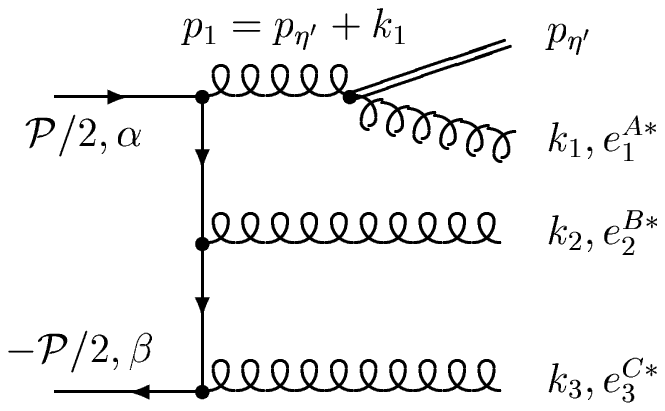}
}}
\caption{A typical Feynman diagram describing the $\Upsilon (1S)
         \to g g g^* (g^* \to \eta^\prime g) \to \eta^\prime X$ decay}
\label{fig:1}
\end{figure}
%
%

In most applications, approximate forms for the
$\eta^\prime$-meson LCDAs are usually employed in which 
only the first non-asymptotic term in both the 
quark-antiquark and gluonic component is kept:
\begin{eqnarray} 
\phi^{(q)}_{\eta^\prime} (x, Q^2) & = & 6 x \bar x
\left [ 1 + 6 (1 - 5 x \bar x) \, A_2 (Q^2) + \ldots \right ] , 
\label{eq:LCDAs} \\ 
\phi^{(g)}_{\eta^\prime} (x, Q^2) & = & 
5 x^2 \bar x^2 \, (x - \bar x) \, B_2 (Q^2) + \ldots~. 
\nonumber 
\end{eqnarray}
These LCDAs involve the Gegenbauer moments for which the following 
notation is used:  
\begin{eqnarray} 
A_2 (Q^2) & = & B^{(q)}_2 \left [ 
\frac{\alpha_s (\mu_0^2)}{\alpha_s (Q^2)} \right ]^{\gamma_+^2}
\!\!\! + \rho_2^{(g)} B^{(g)}_2 \left [
\frac{\alpha_s (\mu_0^2)}{\alpha_s (Q^2)} \right ]^{\gamma_-^2} 
\!\!\! , \quad 
\label{eq:A2} \\
B_2 (Q^2) & = & \rho_2^{(q)} B^{(q)}_2 \left [
\frac{\alpha_s (\mu_0^2)}{\alpha_s (Q^2)} \right ]^{\gamma_+^2}
\!\!\! + B^{(g)}_2 \left [ 
\frac{\alpha_s (\mu_0^2)}{\alpha_s (Q^2)} \right ]^{\gamma_-^2} 
\!\!\! . \quad 
\label{eq:B2} 
\end{eqnarray}
The quantities $\gamma_+^2$, $\gamma_-^2$, $\rho_2^{(q)}$,  
and $\rho_2^{(g)}$ are determined by the perturbative QCD 
while the Gegenbauer coefficients $B^{(q)}_2 (\mu_0^2)$ 
and~$B^{(g)}_2 (\mu_0^2)$ are non-perturbative parameters. 
These coefficients have to be modeled or extracted from a 
phenomenological analysis of experimental data. 

The first attempt to estimate the Gegenbauer coefficients 
$B^{(q)}_2 (\mu_0^2)$ and $B^{(g)}_2(\mu_0^2)$ was recently 
undertaken by Kroll and Passek-Kumericki~\cite{Kroll:2002nt}. 
They performed a NLO theoretical analysis of the 
$\eta^\prime - \gamma$ transition form factor and 
extracted the following values: 
\begin{eqnarray} 
B^{(q)}_2 (\mu_0^2 = 1~{\rm GeV}^2) & = & 0.02 \pm 0.17 , 
\label{eq:GC-KK} \\    
B^{(g)}_2 (\mu_0^2 = 1~{\rm GeV}^2) & = & 9.0  \pm 11.5 ,  
\nonumber 
\end{eqnarray} 
from the CLEO~\cite{Gronberg:1997fj} and
L3~\cite{Acciarri:1997yx} data. 
This fit leaves an order of magnitude uncertainty 
on the coefficients.

The inclusive $\Upsilon (1S) \to \eta^\prime X$ decay   
allows also to get an additional information on the 
Gegenbauer coefficients. 

%
%
%
\begin{figure}[tb]
\centerline{
\resizebox{0.65\hsize}{!}{
\includegraphics*{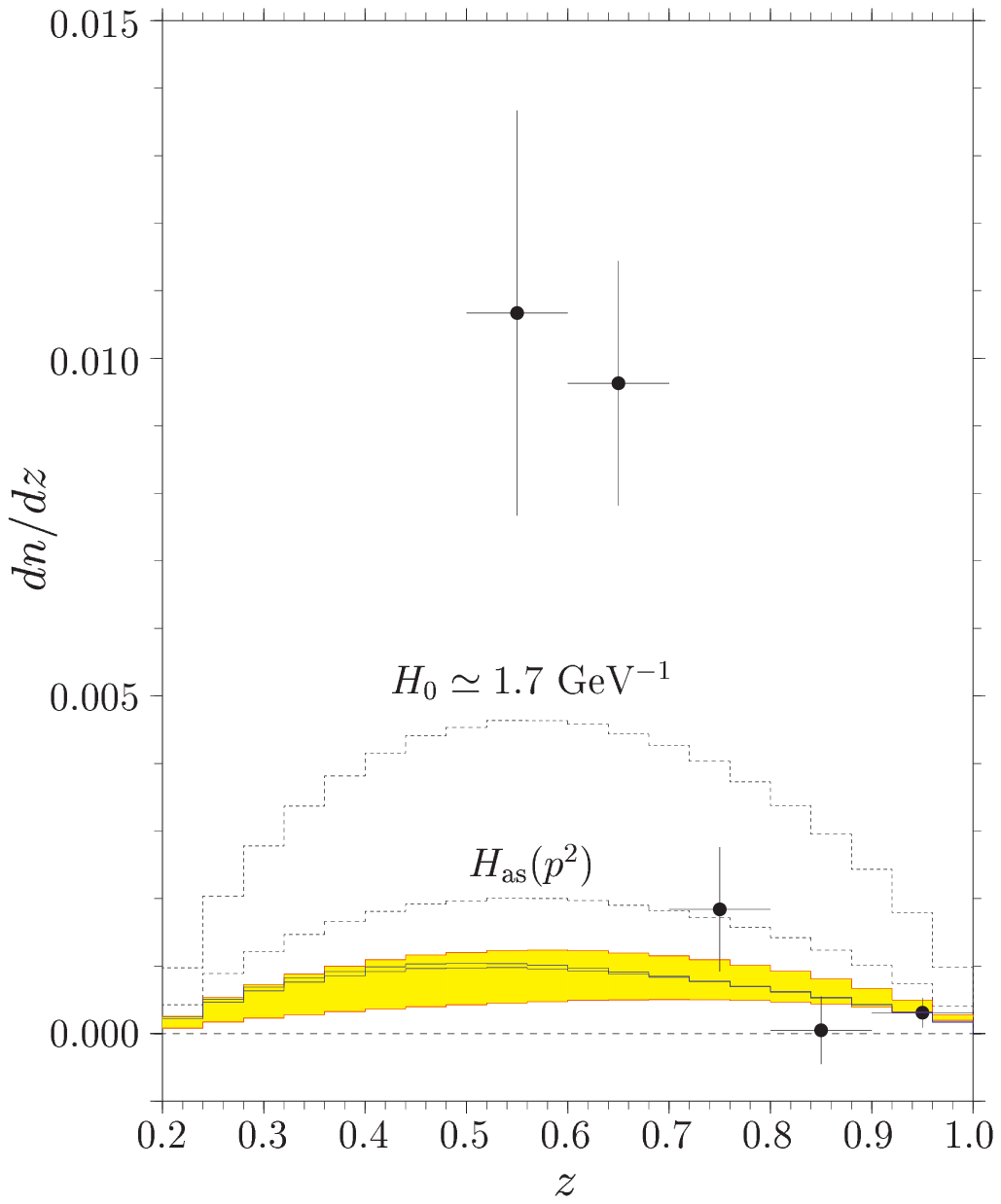}
}}
\caption{The $\eta^\prime$-meson energy spectrum 
         in the $\Upsilon (1S) \to \eta^\prime X$ decay}
\label{fig:8}
\end{figure}
%
%
%
%
\begin{figure}[tb]
\centerline{
\resizebox{0.65\hsize}{!}{
\includegraphics*{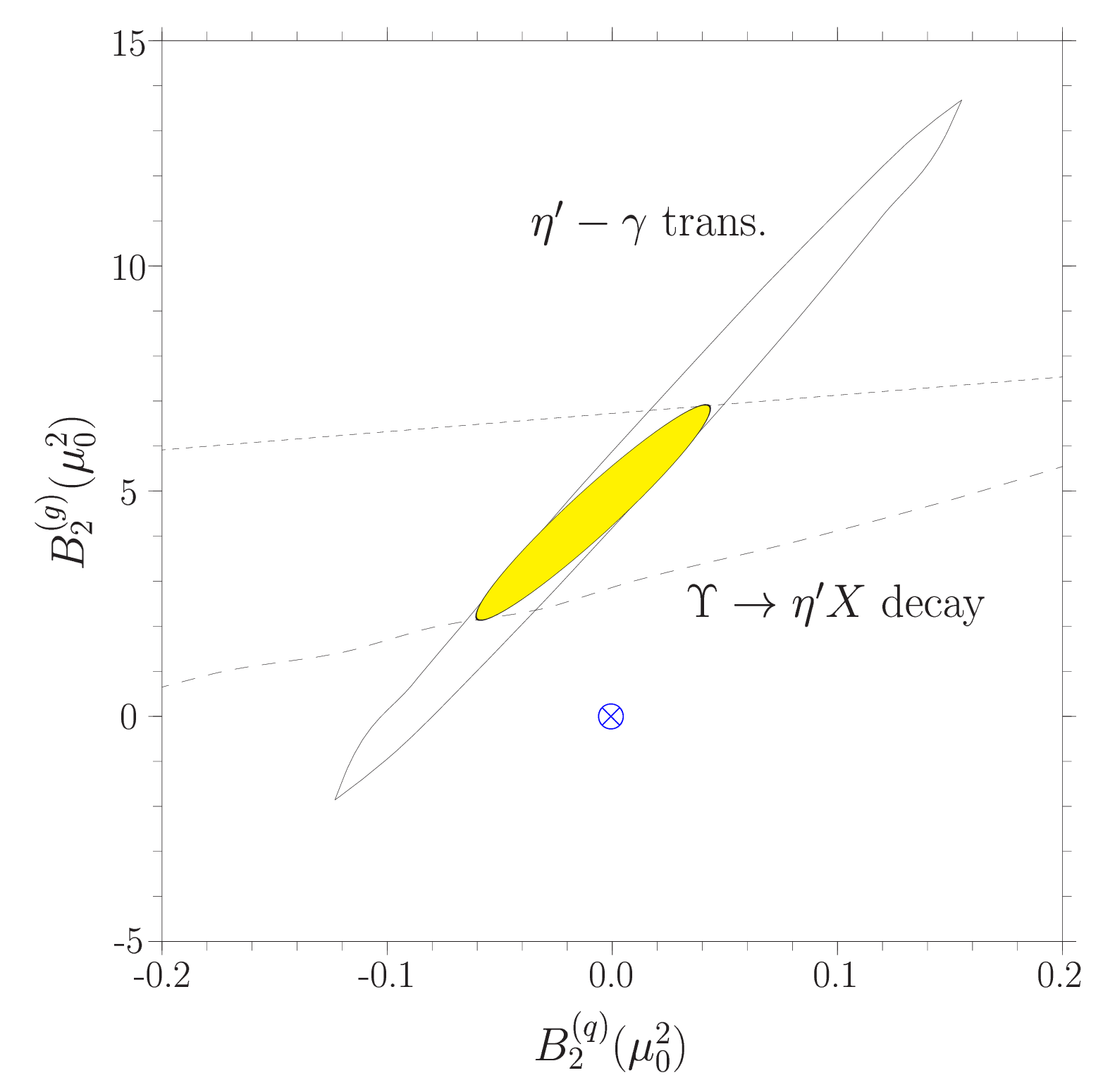}
}}
\caption{
   The resulting 1$\sigma$ contour (combined best fit), shown 
   by the yellow (shaded) region, for the Gegenbauer coefficients 
   estimated at the scale $\mu_0^2=2~{\rm GeV}^2$ from
   the data on the $\eta^\prime - \gamma$ transition form factor
   (solid curve) and $\Upsilon (1S) \to \eta^\prime X$ decay
   (long-dashed and short-dashed curves)}  
\label{fig:5}
\end{figure}
%
\begin{figure*}
\resizebox{0.54\hsize}{!}{
\includegraphics*{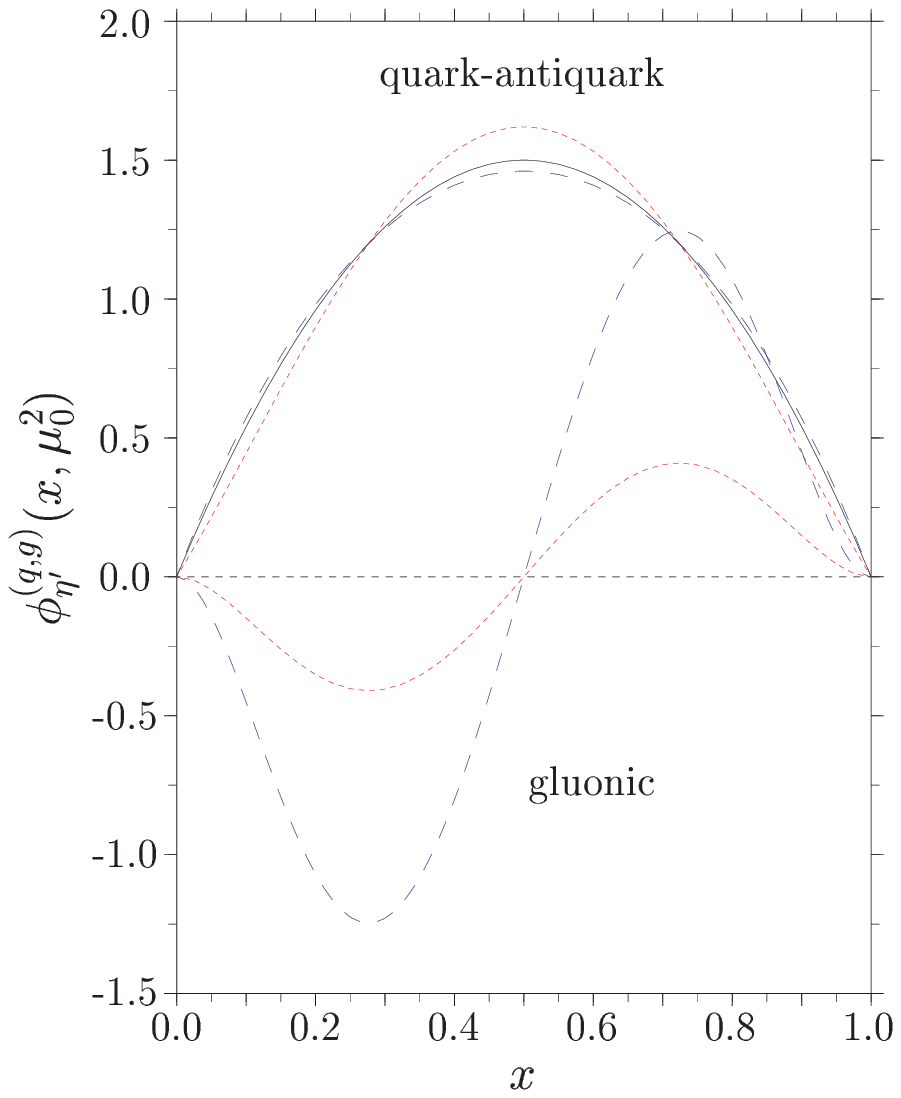}
\includegraphics*{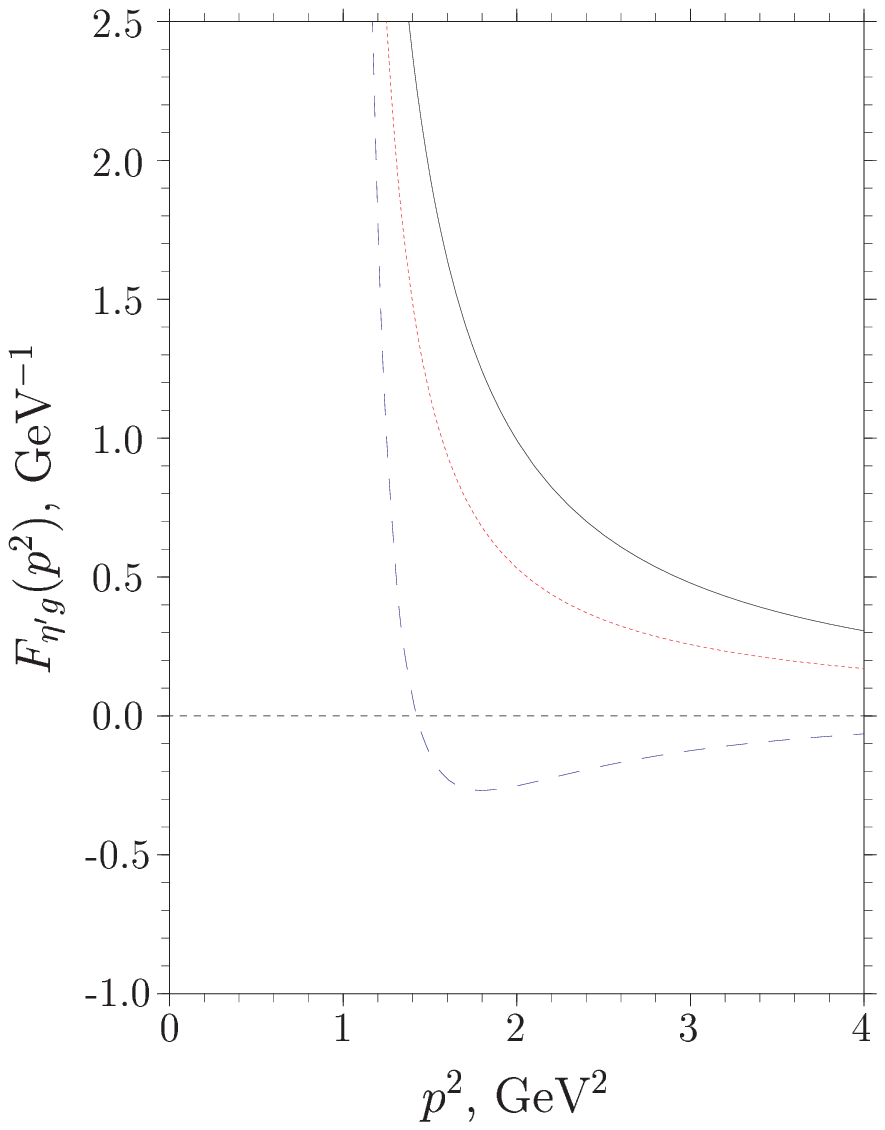}
}
\caption{
      The $\eta^\prime$-meson quark-antiquark $\phi^{(q)}_{\eta^\prime} 
      (x, \mu_0^2)$ and gluonic $\phi^{(g)}_{\eta^\prime} (x, \mu_0^2)$ 
      LCDAs as functions of~$x$ (left frame), and the resulting
      $\eta^\prime - g$ transition form factor (right frame).
      The solid curves correspond to the asymptotic
      quark-antiquark LCDA, while the LCDAs for the central 
      values from the combined best-fit region of the Gegenbauer
      coefficients given in~(\ref{eq:GC-CBF}) are drawn as dotted
      curves. The LCDAs with the values $B^{(q)}_2 = 0.15$
      and $B^{(g)}_2 = 13.5$, which are allowed within~$1\sigma$
      from the analysis of the data on the
      $\eta^\prime - \gamma$ transition form factor~\cite{Kroll:2002nt},
      are presented as the dashed curves. Note that for this case
      the function $F_{\eta^\prime g} (p^2)$ is no longer
      positive definite, as shown in the right frame
}
\label{fig:6}
\end{figure*}
%
%

\section{Perturbative-QCD Analysis of 
         $\Upsilon (1S) \to \eta^\prime X$
         and Comparison with Data}
\label{sec:PQCD-calculations} 

One of the 18~diagrams describing the decay   
$\Upsilon (1S) \to g g g^* (g^* \to \eta^\prime g) 
\to \eta^\prime X$ in the leading order is presented in Fig.~\ref{fig:1};  
the other 17~diagrams can be obtained from the above one by 
the permutations  of the gluons in the intermediate (virtual) 
and final states. The static limit for the heavy quark and 
antiquark in the orthoquarkonium $\Upsilon (1S)$ state 
is used in the calculations. The total decay amplitude 
${\cal M} [\Upsilon \to \eta^\prime g g g]$ is 
rather lengthy and can be found in~\cite{Ali:2003vw}.

The $\eta^\prime$-meson energy spectrum can be theoretically 
determined as follows: 
\begin{eqnarray} 
\frac{dn}{dz} 
& = & \frac{1}{\Gamma_{3g}^{(0)}} \, \frac{1}{3!} \,
\frac{1}{(2 \pi)^8} \, \frac{1}{2 M}
\int \frac{d{\bf k}_1}{2 \omega_1} \, \frac{d{\bf k}_2}{2 \omega_2} \,
\frac{d{\bf k}_3}{2 \omega_3} \,
\frac{d{\bf p_{\eta^\prime}}}{2 E_{\eta^\prime}}
\label{eq:spectrum-def}  \\ 
& \times & 
\delta^{(4)} ({\cal P} - k_1 - k_2 - k_3 - p_{\eta^\prime}) \,
\delta (z - 2 E_{\eta^\prime}/M) 
\nonumber \\ 
& \times &
\frac{1}{3} \sum \left 
|{\cal M} [\Upsilon \to \eta^\prime g g g] \right |^2 ,  
\nonumber 
\end{eqnarray}
where $\Gamma_{3g}^{(0)}$ is the three-gluon decay width of the
$\Upsilon (1S)$-meson in the leading order: 
\begin{equation}
\Gamma_{3 g}^{(0)} = 
\frac{16}{9} \, \left ( \pi^2 - 9 \right ) C_F \, B_F \,
\alpha_s^3 (\mu_\Upsilon^2) \, \frac{|\psi (0)|^2}{M^2} . 
\label{eq:Gamma-3g}
\end{equation}
Here, $C_F = (N_c^2 - 1)/(2 N_c)$, $B_F = (N_c^2 - 4)/(2 N_c)$,
and $\mu_\Upsilon \sim M$ is a typical hard scale of the process. 

As the low-$z$ data are dominated by the fragmentation 
of gluons into the $\eta^\prime$-meson, following the 
CLEO analysis~\cite{Artuso:2002px}, we concentrate on the 
last three ($z \ge 0.7$) and four ($z \ge 0.6$)
experimental bins (see Fig.~\ref{fig:8}). 

The fit of the $B_2^{(q)}(\mu_0^2)$ and $B_2^{(g)}(\mu_0^2)$ 
parameters based on the last four experimental data points results 
in unacceptably large values of the minimum $\chi^2$~\cite{Ali:2003vw}. 
Thus, only the data in the last three bins with $z \ge 0.7$ are used in 
the analysis (quoted $\chi^2$ corresponds to three degrees of freedom) 
and yield the following best fits ($\mu_0^2 = 2$~GeV$^2$): 
\begin{eqnarray} 
B_2^{(q)} (\mu_0^2) & = & -0.89^{+ 1.32}_{- 1.58}, 
\quad 
B_2^{(g)} (\mu_0^2) = -2.86^{+ 20.04}_{- 5.80}, 
\quad 
\label{eq:GC-BF3-1} \\  
B_2^{(q)} (\mu_0^2) & = & -1.09^{+ 1.51}_{- 1.36}, 
\quad 
B_2^{(g)} (\mu_0^2) = 11.53^{+ 5.55}_{- 20.09}, 
\quad 
\label{eq:GC-BF3-2}
\end{eqnarray} 
with $\chi^2 = 2.45$ and~$2.37$ for each set, respectively. 

The overlapping region in the [$B_2^{(q)} (\mu_0^2)$, 
$B_2^{(g)} (\mu_0^2)]$ parameter space from the 
$\eta^\prime - \gamma$ transition form factor~\cite{Kroll:2002nt} 
and the $\Upsilon (1S) \to \eta^\prime X$ decay~\cite{Ali:2003vw} 
is presented in Fig.~\ref{fig:5}. 
The parameter space from the $\Upsilon (1S) \to \eta^\prime X$
decay fit is obtained by imposing the additional condition that
$F_{\eta^\prime g} (p^2)$ remains the positive-definite function 
for $p^2 > m_{\eta^\prime}^2$~\cite{Ali:2003vw} which is 
illustrated in Fig.~\ref{fig:6} in term of both the LCDAs and the 
$\eta^\prime - g$ transition form factor. The resulting combined 
best fit of the Gegenbauer coefficients yields~\cite{Ali:2003vw}: 
\begin{equation}
B_2^{(q)} (\mu_0^2) = -0.008 \pm 0.054 ,
\quad
B_2^{(g)} (\mu_0^2) = 4.6 \pm 2.5 . 
\label{eq:GC-CBF}   
\end{equation}
Using instead the asymptotic LCDA yields $\chi^2 = 8.41$; hence, 
the asymptotic LCDA is not favored by the current analysis. 
However, most of this $\chi^2$ is
contributed by a single experimental point (see Fig.~\ref{fig:8}).  
As data on the hard part of the $\eta^\prime$-meson energy spectrum 
are rather sparse, one can not exclude the asymptotic LCDA based on 
these data. Hopefully, experimental measurements will be improved 
soon to draw more quantitative conclusions.

\section{Summary}
\label{sec:Summary} 

The $\eta^\prime$-meson energy spectrum in the $\Upsilon (1S) \to 
\eta^\prime g g g \to \eta^\prime X$ decay is calculated in the 
leading-order perturbative QCD in the static-quark limit for the 
$\Upsilon (1S)$-meson. The leading-twist (twist-two) quark-antiquark
and gluonic LCDAs are used to describe the $\eta^\prime$-meson 
wave-function. In the LCDAs, the asymptotic and the first 
non-asymp\-to\-tic terms are taken into account. An essential dependence 
of the energy spectrum on the Gegenbauer coefficients is observed. 
These Gegenbauer coefficients are determined in the large-$z$ region 
($z \ge 0.7$) of the $\eta^\prime$-meson energy spectrum from the 
recent CLEO data, however, the resulting $1\sigma$ contour have a 
large dispersion. Combining this analysis with the one of the
$\eta^\prime - \gamma$ transition form factor and requiring 
additionally that the EVF, $F_{\eta^\prime g} (p^2)$,  remains 
positive-definite in the entire $p^2 > m_{\eta^\prime}^2$ region, 
yield much improved determination of the Gegenbauer coefficients.   

It remains to be seen if the so-determined $\eta^\prime g^* g^*$ EVF  
explains the data on the inclusive $B \to \eta^\prime X_s$ decay.  

\begin{acknowledgement}

The work of A.Ya.P. has been supported by the 
Schweizerischer Nationalfonds.

\end{acknowledgement}

\end{document}